\documentclass[secnumarabic, graphics,floatfix,tightenlines,nobibnotes, aps, prb, twocolumn]{revtex4-1}
\usepackage{amsmath,amssymb}
\usepackage{graphicx}% Include figure files
\usepackage{dcolumn}% Align table columns on decimal point
\usepackage{bm}% bold math
\usepackage{braket}
\usepackage{verbatim}
\usepackage{float}
\usepackage{bbold}
\usepackage{color}
\usepackage{xcolor}
\usepackage{relsize}

\usepackage{slashed}
\usepackage{amsthm}
\usepackage[colorlinks=true ,urlcolor=blue,urlbordercolor={0 1 1}]{hyperref}
\begin{document}

\title{Non-Abelian and Abelian descendants of vortex spin liquid: fractional quantum spin Hall effect in twisted MoTe$_2$}
\author{%
Ya-Hui Zhang
}
\affiliation{William H. Miller III Department of Physics and Astronomy, Johns Hopkins University, Baltimore, Maryland, 21218, USA}

\date{\today}% It is always \today, today,

\begin{abstract}
The recent experimental observation of a potential fractional quantum spin Hall (FQSH) state in the twisted MoTe$_2$ system has sparked theoretical explorations  at total filling $\nu_T=1$ of a pair of $C=\pm 1$ Chern bands from the two spins (locked to valley). One intriguing candidate is a vortex spin liquid (VSL), which can be viewed as an exciton version of composite fermi liquid (CFL). The VSL insulator is incompressible in the charge channel, while compressible in the spin channel. Here we investigate fully gapped descendants of the VSL phase at the total filling $\nu_T=\nu_{\uparrow}+\nu_{\downarrow}=1$. At zero magnetization,  a non-Abelian state with both FQSH and thermal Hall effect emerges from weak $p+ip$ pairing of the neutral Fermi surface, hosting 12 anyons (up to addition of physical electron) including two independent Ising anyons separately carrying charge and spin.   Strong pairing leads to a Z$_4$ topological order with only FQSH effect. At non-zero magnetization $m=2S_z$, there is a Jain sequence of magnetic plateaus with $m=\frac{1}{p}, p \in Z$, exhibiting both half FQSH effect and spin fractional quantum Hall effect (SFQH).   Our work highlights the VSL's potential  as a parent state to organize numerous FQSH insulators with non-trivial inter-valley correlations at  $\nu_T=1$. The quantized FQSH behavior remains robust even in the presence of ferromagnetism, thanks to the spin-charge separation nature inherent in the parent VSL phase. Future experimental investigations are crucial to validate or rule out spontaneous magnetization and time reversal symmetry breaking.
\end{abstract}

\pacs{Valid PACS appear here}% PACS, the Physics and Astronomy
\maketitle

\section{Introduction}

Strong correlated physics in the moir\'e Chern bands\cite{zhang2019nearly,wu2019topological} have recently attracted many attentions. In the moir\'e system one can find a pair of Chern bands with opposite Chern numbers in the two valleys.  Due to strong Coulomb exchange, fully valley polarized state is often favored\cite{zhang2019nearly,repellin2020ferromagnetism}, leading to a single Chern band similar to the Landau level at high magnetic field. Therefore we can expect these moir\'e chern bands\cite{zhang2019nearly} to be wonderful platforms  for fractional quantum anomalous Hall effect(FQAH) from fractional Chern insulator(FCI) at zero magnetic field\cite{sun2011nearly,sheng2011fractional,neupert2011fractional,wang2011fractional,tang2011high,regnault2011fractional,bergholtz2013topological,parameswaran2013fractional,PhysRevB.84.165107}. Indeed FQAH states have been observed in twisted MoTe$_2$ homobilayer \cite{cai2023signatures,zeng2023integer,park2023observation,PhysRevX.13.031037} and in   rhombohedrally stacked multilayer graphene aligned with hBN\cite{2023arXiv230917436L}, which also attracted many theoretical investigations for the twisted MoTe$_2$ system\cite{wu2019topological,yu2020giant,devakul2021magic,li2021spontaneous,crepel2023fci,wang2023fractional,reddy2023fractional,2023arXiv230809697X,2023arXiv230914429Y,PhysRevLett.131.136501,PhysRevLett.131.136502,morales2023magic,song2023phase} and the pentalayer graphene system\cite{dong2023theory,zhou2023fractional,dong2023anomalous,guo2023theory,kwan2023moir}.

One interesting question is whether new physics can be realized different from the familiar quantum Hall systems. One unique feature of the moir\'e system is that the Chern bands always exist as a pair with opposite Chern numbers due to the time reversal symmetry. For example, in the twisted MoTe$_2$ system, one always finds a $C=1$ Chern band in one valley and a conjugate $C=-1$ Chern band in the other valley. Therefore we can simulate a completely new Coulomb coupled quantum Hall bilayer with opposite chiralities in the two layers by treating the valley as a `layer'. The author already considered the total filling $\nu_T=\frac{1}{2}+\frac{1}{2}$ case of this conjugate bilayer in a previous paper in 2018\cite{zhang2018composite} and predicted a new incompressible state called composite fermion insulator. Such a state has a neutral Fermi surface and gapless exciton excitations.  

More recently,  a state with vanishing charge Hall conductivity and fractional quantum spin Hall (FQSH) effect was reported experimentally in twisted MoTe$_2$ at the hole filling $\nu_T=3$ with twist angle $\theta=2.1^\circ$\cite{kang2023fqsh}.  In this state, the first Chern band in both valleys is fully occupied and we can consider the total filling $\nu_T=1$ of the second Chern band.  This is exactly the conjugate Chern band case considered by the author in Ref.~\onlinecite{zhang2018composite}.  Actually, the previously proposed composite fermion insulator\cite{zhang2018composite} also exhibits quantized FQSH and is a strong candidate for the FQSH state observed in the experiment\cite{zhang2024vortex}. In the context of the twisted MoTe$_2$ system, as the spin is locked to valley, the internal flux of this composite fermion insulator carries gapless spin excitation and we dub the phase as vortex spin liquid (VSL).

\begin{figure*}[hbt]
    \centering
    \includegraphics[width=0.95\textwidth]{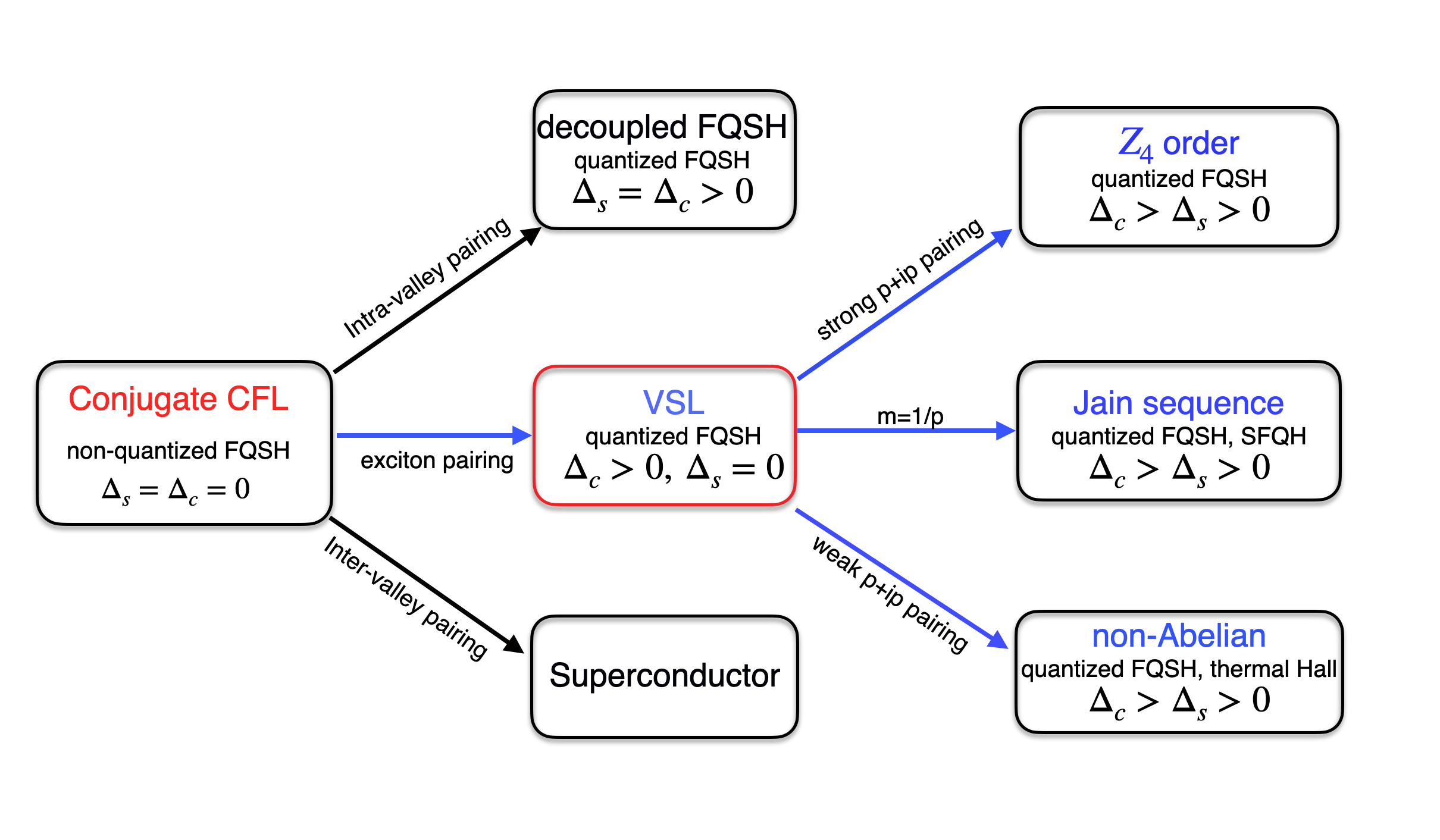}
    \caption{The relationship of different descendants of the conjugate composite Fermi liquid (cCFL) at total filling $\nu_T=\nu_{\uparrow}+\nu_{\downarrow}=1$. We can reach various phases from Cooper pairing or exciton pairing of the composite fermions. Due to inter-valley Coulomb repulsion, we postulate the vortex spin liquid (VSL) from exciton pairing to be favored among the second generation. Below the charge gap, the VSL phase is an exciton version of composite Fermi liquid (CFL) and is a parent state for the descendants at different exciton density decided by the magnetizaiton $m=2\langle S_z \rangle$. At zero magnetization, weak and strong p+ip pairing leads to a non-Abelian topological order with 12 anyons and an Abelian $Z_4$ topological order. The non-Abelian pfaffian-like state also has a quantized thermal Hall effect with chiral central charge $c=\frac{1}{2}$. At non-zero $m=\frac{1}{p}, p\in Z$, there emerges a Jain sequence of maganetic plateaus with both quantized fractional quantum spin Hall effect(FQSH) with $\frac{1}{4\pi}A_c d A_s$ response and spin fractional quantum Hall effect (SFQH) with $\frac{1}{4p} \frac{1}{4\pi}A_s d A_s$ response. It also hosts thermal Hall effect with chiral central charge $c=-(p-1)$.  }
    \label{fig:vsl_descendants}
\end{figure*}

VSL can be constructed from inter-valley exciton pairing\cite{zhang2018composite,zhang2024vortex,shi2024excitonic} of the composite fermions (CF) from a decoupled composite Fermi liquids at half filling of the conjugate $C=\pm 1$ Chern bands\cite{zhang2018composite,myerson2023conjugate}. It can be viewed as an exciton version of the composite Fermi liquid (CFL)\cite{halperin1993theory,son2015composite} with also particle-hole symmetry. The VSL has a bulk charge gap with gapless helical edge modes.  The gapless bulk excitations are purely neutral and should not influence any charge transport measurements.  In the familiar Landau level system, it is known that the CFL phase is a parent state for gapped topological orders at half-filling or fillings doped away from half filling\cite{jain2000composite,jain2007composite,jain1989composite}. Following the same spirit, in this paper we study possible fully gapped descendants of the VSL phase at zero magnetization or non-zero magnetization $m=2\langle S_z \rangle$ (see Fig.~\ref{fig:vsl_descendants}). At zero $m$, we can obtain a gapped state from $p+ip$ pairing of the neutral Fermi surface in the VSL phase. In the weak pairing case, this leads to a non-Abelian state with 12 anyons (up to addition of the physical electron), including two independent Ising anyons carrying charge and spin separately. The non-Abelian state also has a quantized thermal Hall effect. In the strong pairing case we obtain a Z$_4$ topological order with only FQSH effect.  At finite magnetization, there is a Jain sequence of Abelian state at $m=\frac{1}{p},\ p\in Z$, where the neutral fermion forms a Chern insulator with $C=-p$. These states exhibit both quantized half FQSH and a quantized spin fractional quantum Hall effect(SFQH). Note SFQH is defined as a spin-spin Hall conductivity and can not be accessed in any charge transport measurements. All of these descendants share the property of the half quantized FQSH and spin-charge separation from the parent VSL state and they are all consistent with the current experiment\cite{kang2023fqsh}. We suggest future experiments to measure the valley polarization and time reversal symmetry breaking to further distinguish them.

\section{Low energy field theory of the VSL phase}

We can construct\cite{zhang2018composite,zhang2024vortex} the VSL phase from inter-valley exciton pairing $\Phi f^\dagger_+ f_{-}$ of the composite fermion $f_\pm$ from the conjugate CFL. Here $\pm$ labels the two valleys with $C=\pm 1$.  Depending on the value of $\Phi$ and also lattice superlattice potential, there are many possibilities for the neutral fermi surface formed by $f_+\sim f_- \sim f$. Here we assume a single Fermi surface with volume $1$ in units of the moire BZ. We ignore the moir\'e  superlattice potential which is appropriate in the Landau level limit.

The low energy field theory of the VSL phase is\cite{zhang2024vortex}:

\begin{align}
\mathcal L_{\text{VSL}}&=\mathcal L_{FS}[f,a]-\frac{1}{4\pi}A_s da+\frac{1}{4\pi}A_c d A_s
\label{eq:VSL}
\end{align}
with
\begin{align}
    \mathcal L_{\text{FS}}[f,a]&=f^\dagger(t,\mathbf x)(i\partial_t+a_0)f(t,\mathbf x)\notag \\ 
    &~~~-\frac{\hbar^2}{2m} f^\dagger(t,\mathbf x)(-i\partial_\mu-a_\mu)^2f(t,\mathbf x)
\end{align}

Here $a_\mu$ is an internal U(1) gauge field. $A_{c;\mu}=\frac{1}{2}(A_{+;\mu}+A_{-;\mu})$ and $A_{s;\mu}=A_{+;\mu}-A_{-;\mu}$ are the probing U(1) gauge fields in the charge and spin channel.  $A_{\pm ;\mu}$ is the U(1) probing field coupled to the valley $\tau=\pm$ which has Chern number $C=\pm$ respectively.  $adb$ is an abbreviation of the Chern-Simons term $\epsilon_{\mu \nu \sigma} a_\mu \partial_\nu b_\sigma$ with $\epsilon$ the anti-symmetric tensor.

The above action resembles Son's Dirac theory\cite{son2015composite} of the CFL in half filled Landau level. Actually the VSL phase can be viewed as a CFL for the neutral excitons charged under $A_s$.  The microscopic time reversal symmetry $\mathcal T$ acts as an effective particle-hole symmetry CT for the excitons and forbids the interal Chern-Simons term. For bosonic excitons CT$^2=1$ and the neutral Fermi surface can be from ordinary dispersion instead of Dirac fermion.

In the following we will consider the fully gapped topological orders as descendants of the VSL phase.   The above action loses the information of the gapped excitations carrying the charge under $A_c$, which is encoded in the vortex of the composite fermion exciton order $\Phi$. For the purpose to obtain the full data of the topological order, we need to consider the  complete action:

\begin{align}
    \mathcal L_{\text{Full-VSL}}&=\mathcal L_{\text{FS}}[f,\frac{a_++a_-}{2}]+\frac{1}{2\pi}b_1 d (a_+-a_-)\notag \\
    &~~~-\frac{2}{4\pi}\alpha_+ d \alpha_++\frac{1}{2\pi}(A_+-a_+)d\alpha_+ \notag \\
    &~~~+\frac{2}{4\pi}\alpha_- d \alpha_-+\frac{1}{2\pi}(A_--a_-)d\alpha_- 
    \label{eq:full_vsl}
\end{align}

In the first line, we introduce another U(1) gauge field $b_{1;\mu}$ to represent the CF exciton condensation order $\Phi$. The charge of $b_1$ represents the vortex of $\Phi$ and carries the physical charge. $\alpha_\pm$ are introduced to provide the Chern-Simons term in the conjugate CFL\cite{zhang2024vortex}. If we do not care about the gapped charge excitations, we can integrate $b_1$ which locks $a_+=a_-=a$. Then integration of $\alpha_\pm$ gives the term $-\frac{1}{4\pi}A_s d a+\frac{1}{4\pi} A_c d A_s$.

\section{Non-Abelian FQSH insulator at zero valley polarization}

We consider a $p+ip$ pairing of the neutral Fermi surface $f$. The resulting phase is a non-Abelian topological order with 12 anyons in the weak pairing case and a Z$_4$ topological order in the strong pairing case.

\subsection{Strong pairing limit: Z$_4$ topological order}
First  let us consider the strong pairing limit so we do not need to worry about the Majorana mode.
As we already have an exciton order $\Phi f^\dagger_+ f_{-}$, we now introduce the pairing term $\Delta(\mathbf k) f(\mathbf k) f(-\mathbf k) \sim \Delta(\mathbf k) f_{+}(\mathbf k) f_{-}(-\mathbf k)$. Starting from the full action of the VSL in Eq.~\ref{eq:full_vsl}, we introduce a U(1) gauge field $b_2$ as the dual field of $\Delta$, the resulting theory is:
\begin{align}
    \mathcal L&=\frac{1}{2\pi} b_1 d(a_+-a_-)+\frac{1}{2\pi}b_2 d(a_++a_-) \notag \\ 
    &~~~-\frac{2}{4\pi}\alpha_+ d \alpha_++\frac{2}{4\pi}\alpha_{-}d\alpha_{-} \notag \\ 
&~~~+\frac{1}{2\pi} (A_+-a_+) d \alpha_++\frac{1}{2\pi}(A_--a_-)d\alpha_-
\end{align}

Integration of $a_\pm$ locks $\alpha_+=b_1+b_2$ and $\alpha_-=b_2-b_1$. The final action is:

\begin{align}
    \mathcal L_{Z_4}=-\frac{4}{2\pi}b_1 d b_2+\frac{2}{2\pi}A_c d b_2+\frac{1}{2\pi} A_s d b_1
    \label{eq:Z4}
\end{align}

This is a Z$_4$ topological order with K matrix $K=\begin{pmatrix} 0 & 4\\ 4 & 0 \end{pmatrix}$.  We have the charge vector $q_c=(0,2)^T$ for $A_c$ and  $q_s=(1,0)^T$ for $A_s$.  It is easy to get $\sigma_{cs;xy}=\frac{1}{2} $ while $\sigma_{cc;xy}=\sigma_{ss;xy}=0$.   We have two minimal anyons: (1) a boson generated by $l_1=(1,0)^T$ with charge $Q=\frac{1}{2}, S_z=0$. This corresponds to the vortex of the exciton order $\Phi$.  (2)  a boson generated by $l_2=(0,1)^T$ with charge $Q=0, S_z=\frac{1}{4}$. This corresponds to the vortex of the cooper pairing order $\Delta$.   Therefore we have spin-charge separation in this topological order as is inherited from the parent VSL phase. We expect that the spin gap is smaller than the charge gap as the later is decided by the Coulomb interaction scale while the former is from spin interaction.

The edge theory of the Z$_4$ topological order is:

\begin{align}
    S&=\int dt dx \frac{4}{2\pi} \partial_x \varphi_1 \partial_t \varphi_2-\upsilon (\partial_x \varphi_1)^2-\upsilon(\partial_x \varphi_2)^2 \notag \\ 
    &~~~-g\partial_x \varphi_1 \partial_x \varphi_2
    \label{eq:Z4_edge}
\end{align}

One can see that it is in the form of the usual Luttinger liquid.  But now $e^{i \varphi_1}$ creates an excitation with $Q=\frac{1}{2}, S_z=0$ and $e^{i \varphi_2}$ creates an excitation with $Q=0, S_z=\frac{1}{4}$.  $e^{i 2 (\varphi_1+\varphi_2)}$ creates an excitation with $Q=1,S_z=1/2$, similar to physical electron. However it is a bosonic field and should not be identified with the physical electron operator.  The Z$_4$ topological order is essentially a bosonic state with single electron trivially gapped.  As a result, there is still single electron gap at the edge.

\subsection{Weak pairing limit: non-abelian order with $12$ anyons}

In the weak pairing limit, the vortex of the cooper pairing hosts a majorana zero mode\cite{read2000paired,moore1991nonabelions}. Actually the vortex of the exciton order $\Phi$ also has a majorana zero mode.   In the vortex core of $\Phi$, $\Phi$ vanishes and we have two fermi surfaces from $f_+, f_-$. Then the $p+ip$ pairing of $\Delta(\mathbf k) f_+(\mathbf k) f_-(-\mathbf k)$ gives a Chern number $C=2$, while the bulk has $C=1$ in the large $\Phi$ region.  As a consequence, there is a domain wall separating $C=2$ and $C=1$ around the vortex of the $\Phi$ order, leading to a majorana zero mode.

 The final topological order is (Z$_4$ $\times $ Ising)/$Z_2$, where the Z$_4$ corresponds to the action in Eq.~\ref{eq:Z4}. Ising means the topological order with three anyons $I, \psi, \sigma$ and fusion rule: $\sigma \times \sigma=I+\psi$, $\psi\times \psi=I$, $\psi \times \sigma=\sigma$.  $I$ is the trivial anyon, $\psi$ is a neutral fermion (the bogoliubov quasiparticle) and $\sigma$ is the Ising anyon (bound to the elementary vortex of $\Phi$ or $\Delta$).  There is a $\pi$ mutual statistics between the fermion $\psi$ and the Ising anyon $\sigma$. $\sigma$ has self statistics $\theta_{\sigma}=e^{i\frac{\pi}{8}}$.   

 We can label the anyons of this topological order as $(l_1,l_2, I)$, $(l_1,l_2,\psi)$ and $(l_1,l_2,\sigma)$. $(l_1,l_2)$ are the charges under $b_1, b_2$ in the Z$_4$ topological order in Eq.~\ref{eq:Z4}. $(l_1,l_2)$ corresponds to $l_1$-vortex of $\Phi$ and $l_2$-vortex of $\Delta$. The majorana mode (the $\sigma$ anyon) is bound to odd $l_1+l_2$ and the $1,\psi$ anyons are bound to even $l_1+l_2$.   Then there are only $(4\times 4 \times 3)/2=24$ choices.  Besides, we also find that $(2,2,I), (2,2,\psi)$ are equivalent to $I=(0,0,I),\psi=(0,0,\psi)$ up to addition of the physical electron.   It is easy to find that $(2,2,I)$ and $(2,2,\psi)$ have charges $(Q,S_z)=(1,\frac{1}{2})$ simply from the Z$_4$ part. They have self statistics $0, \pi$ respectively. $(2,2,\psi)$ has trivial mutual statistics with all of the other anyons and  we can identify $(2,2,\psi)=e$ as the physical electron. Then $(2,2,I)= e \psi$. Given that we are considering a fermionic system formed by electrons, $e$ should be viewed as a trivial particle: $e \simeq I$. Then we have $(2,2,\psi)\simeq (0,0, I)$ and $(2,2,I) \simeq (0,0,\psi)$.   As a result, $(l_1,l_2)$ is equivalent to $(l_1+2,l_2+2)$, which further reduces the number of anyons to  12.  To our best knowledge, this is the minimal topological order with smallest ground state degeneracy on torus consistent with the quantized half FQSH, zero charge Hall conductivity and zero magnetization. In contrast, if we consider the decoupled FQSH insulator with a Pfaffian state in one valley and its conjugate in the other valley, the number of anyons is $6 \times 6=36$ and is much larger.

Among the 12 anyons, there are two elementary Ising anyons: $\sigma_1=(1,0,\sigma)$ and $\sigma_2=(0,1,\sigma)$. One can find that $\sigma_1$ has charge $(Q,S_z)=(\frac{1}{2},0)$ and $\sigma_2$ has charge $(Q,S_z)=(0,\frac{1}{4})$. Thus we have spin-charge separation: the gap of $\sigma_1$, $\sigma_2$ set the charge gap and spin gap respectively.  We again expect the spin gap to be smaller than the charge gap.

This non-Abelian state breaks time reversal symmetry and has a chiral central charge $c=\frac{1}{2}$ and thus thermal Hall effect. The edge theory is the same theory in Eq.~\ref{eq:Z4_edge} with an additional chiral Ising conformal field theory (CFT).  In this time we have a physical electron operator as $c^\dagger_\uparrow \sim \psi e^{i 2 (\varphi_1+\varphi_2)}$ and thus there is no single electron gap at the edge.

Lastly we comment on the difference between the strong pairing and weak pairing state.   Their relationship is quite similar to that between the strong pairing $U(1)_8$ Abelian order and weak pairing Pfaffian state from the CFL in half-filled Landau level. In the previous numerical studies of the half-filled Landau level, the weak pairing seems to be more likely than the strong pairing state.   Therefore, for our case, we also expect the non-Abelian state to be more relevant than the Z$_4$ state.  The non-Abelian state necessarily breaks the time reversal symmetry.  But the detection of the time reversal symmetry breaking is subtle because the charge Hall effect and valley polarization both vanish. The state is definitely compatible with the current experiment and may be considered seriously in future experimental investigations.

\section{Jain sequence at non-zero magnetization $m=\frac{1}{p}$: quantized FQSH and SFQH }

Now we move to the non-zero magnetization corresponding to the filling $\nu_{\uparrow}=\frac{1}{2}+\frac{m}{2}$ and $\nu_{\downarrow}=\frac{1}{2}-\frac{m}{2}$. This is equivalent to tuning the density of the excitons and we expect a Jain sequence of gapped states at different magnetization $m=\frac{1}{p}, p \in Z$.

First, from  action of the VSL in Eq.~\ref{eq:VSL}, we have the following relationship between the internal magnetic field $b=\partial_x a_y-\partial_y a_x$ and the magnetization: $\frac{b}{2\pi}=-m$.  Thus at non-zero magnetization $m$, the neutral fermion $f$ feels an effective flux and forms Landau levels. Its effective filling is at $\nu=\frac{n_f}{b/(2\pi)}=-\frac{1}{m}$ where we used $n_f=1$ per moir\'e unit cell.  Therefore the neutral fermion $f$ is in an integer quantum Hall state with $C=-p$ when $m=\frac{1}{p}$, $p=2,3,4,...$.  Note that $p=1$ is impossible a we must have $m<1$.  At $m=\frac{1}{p}$, there is a self Chern-Simons term $-\frac{p}{4\pi} a d a$.  Integration of $a$ gives the final response $\mathcal L_{\text{response}}=\frac{1}{4p} \frac{1}{4\pi} A_s d A_s +\frac{1}{4\pi} A_c d A_s$.  Hence these gapped states have both spin fractional quantum Hall effect (SFQH) with $\sigma_{ss;xy}=\frac{1}{4p}$ and the FQSH with $\sigma_{cs;xy}=\frac{1}{2}$.   There is also quantized thermal Hall effect corresponding to a chiral central charge $c=-(p-1)$. Note that the $C=-p$ Chern insulator gives $p$ number of edge modes, but one will be removed by the coupling to the gauge field $a_\mu$. This will become clearer in the following K matrix description.

To obtain the precise data of the Abelian topological order, we should start from the full action of the VSL phase in Eq.~\ref{eq:full_vsl}.  We introduce $p$ number of U(1) gauge fields $\beta_I$ to capture the $C=-p$ integer quantum Hall state of the neutral fermion $f$. The final action is:

\begin{align}
    \mathcal L&=\sum_{I=1}^p \frac{1}{4\pi} \beta_I d \beta_I + \frac{1}{2\pi} \sum_{I=1}^p \frac{a_++a_-}{2} d \beta_I+\frac{1}{2\pi} b_1 d (a_+-a_-) \notag \\ 
    &~~~-\frac{2}{4\pi} \alpha_+ d \alpha_++\frac{1}{2\pi} (A_{+}-a_+) d \alpha_+ \notag \\ 
    &~~~+\frac{2}{4\pi}\alpha_- d \alpha_-+\frac{1}{2\pi} (A_--a_-) d \alpha_-
\end{align}

We can integrate $b_1$ to lock $a_+=a_-=a$. Then integration of $a$ leads to the constraint $\alpha_++\alpha_-=\sum_{I=1}^{p} \beta_I$.   In the end we can use $\beta_p=\alpha_++\alpha_--\sum_{I=1}^{p-1}\beta_I$ to rewrite the action.  This leads to a K matrix with dimension $(p+1) \times (p+1)$.   In the order of $(\alpha_+,\alpha_-,\beta_1,...,\beta_{p-1}$, the $K$ matrix is $K=\begin{pmatrix} 1 & -1 & 1 \\ -1 & -3 &1 \\ 1 &1 &-2 \end{pmatrix}$ for $p=2$ and $K=\begin{pmatrix} 1& -1 & 1 &1 \\ -1 & -3 & 1 &1 \\ 1 & 1 & -2 & -1 \\ 1& 1 & -1 &-2 \end{pmatrix}$ for $p=3$.   Larger $p$ can be easily generalized: the left top $2\times 2$ block remains the same; the right bottom $(p-1) \times (p-1)$ block has $-2$ in the diagonal and $-1$ in the other entry.   Other entries in the right-top and left-bottom $2\times (p-1)$ block are all $1$.  The charge vector is $q_c=(1,1,0,...,0)^T$ under $A_c$ and $q_s=(\frac{1}{2},-\frac{1}{2},0,...,0)^T$.  It may also be interesting to note that the right-bottom $(p-1)\times (p-1)$ block is the same as the K matrix of the $SU(p)_1$ chiral spin liquid.  One can verify that the chiral central charge $c=-(p-1)$.

There is a more simplified form to obtain the charge and statistics of the anyon. We can simply integrate all of $\beta_I$ field and reach an action in terms of $(\alpha_+, \alpha_-,a)$. In this approach we lose the information of the thermal Hall effect and also we need to keep in mind that the self statistics from charge under $a$ should be shifted by $\pi$ because the elementary charge $f$ is a fermion.  The $K$ matrix is $K=\begin{pmatrix} 2 & 0 & 1 \\ 0 & -2 & 1 \\ 1 & 1 & p \end{pmatrix}$ in the order of $(\alpha_+,\alpha_-,a)$. We have charge vector $q_c=(1,1,0)^T$ and $q_s=(\frac{1}{2},-\frac{1}{2},0)^T$. We find $K^{-1}=\frac{1}{4p} \begin{pmatrix} 2p+1 & -1 & -2 \\ -1 & 1-2p & 2 \\ -2 & 2 &4\end{pmatrix}$.  There are $4p$ number of anyons because $\det K=-4p$. It is easy to find that $l=(0,0,-1)^T$ generates an anyon with charge $(Q,S_z)=(0,\frac{1}{2p})$. It has statistics $\theta=\frac{\pi}{p}-\pi$.  $l=(1,0,0)^T$ generates an anyon with charge $(Q,S_z)=(\frac{1}{2},\frac{1+p}{4p})$ and statistics $\theta=\frac{2p+1}{4p}\pi$. Similarly $l=(0,-1,0)^T$ creates an anyon with charge $(Q,S_z)=(\frac{1}{2},\frac{1-p}{4p})$ and statistics $\theta=\frac{1-2p}{4p} \pi$.

Just from the Hall response we can read that there is an anyon with charge $Q=\frac{1}{2}, S_z=\frac{1\pm p}{4p}$ which can be generated from threading a $d A_c=\pi, d A_s=\pm 2\pi $ flux (for example, from $dA_+=2\pi$, $d A_-=0$).  Meanwhile from threading a flux of $d A_c=2\pi, d A_s=0$ one can create a different anyon with $Q=0, S_z=\frac{1}{2}$. The above analysis from the K matrix actually  shows that there is a neutral anyon with even smaller $S_z=\frac{1}{2p}$. We expect this neutral anyon to be energetically cheaper than the charged anyon and sets a spin gap smaller than the charge gap.

\section{Discussion on experiment}

In this paper we show that there are many descendant fully gapped topological orders from the vortex spin liquid in the conjugate moir\'e Chern band at the total filling $\nu_T=1$. In the current experiment\cite{kang2023fqsh}, we only know that there is a half FQSH effect and the charge Hall conductivity vanishes. This is actually consistent with the VSL phase and all of its descendants discussed here. Actually it is also consistent with the more naive decoupled phase with one quantum Hall state in one valley and its conjugate partner in the other valley.  Compared to the decoupled phase, the VSL phase and its descendants all have the property of spin charge separation. We expect a finite charge gap $\Delta_c$ set by the Coulomb interaction and a much smaller or even zero spin gap $\Delta_s$.  In contrast, in the decoupled phase $\Delta_c=\Delta_s$.

Another important observation is that the quantized half FQSH is compatible with time reversal symmetry breaking and even a finite valley polarization $m=2\langle S_z \rangle$. This is unprecedented in the quantum spin Hall band insulator where a ferromagnetism of $S_z$ would make the system metallic. Due to the spin-charge separation nature of the VSL phase, one can deform the spin part and change the magnetization continuously without influencing the charge transport and the helical edge modes.  We suggest the future experiment to measure the magnetization $m$ at zero magnetic field $B$ through magnetic circular dichroism (MCD).  If $m=\frac{1}{p}, p \in Z$ at $B=0$, then it may likely be in one of the Jain sequence state with both quantized FQSH and spin fractional quantum Hall effect (SFQH) with $\sigma_{ss;xy}=\frac{1}{4p}$. If $m$ is finite but not equal to $\frac{1}{p}$, it may be in a gapless state with charge gap, but zero spin gap. In this case it still has quantized FQSH, but a non-quantized SFQH. Even if $m=0$ when $B=0$, one can push the system into the state with non-zero $m$  by applying a Zeeman field. Observing quantized FQSH with the same charge transport behavior under varying magnetization would be a strong evidence of spin-charge separation. Actually the existence of the Jain sequence at $m=\frac{1}{p}$ suggests quantum oscillations under the Zeeman field.  We expect certain quantity like the density of states oscillates in the form $\cos (2\pi p)=\cos (\frac{2\pi}{m}) \sim \cos ( \frac{2\pi}{\chi_s g B})$ with $\chi_s$ the spin susceptibility. Thus there is a quantum oscillation in terms of $\frac{1}{B}$, but now with a potentially very different coefficient from the usual quantum oscillation of a Fermi liquid under the orbital magnetic field.

If a FQSH insulator with zero magnetization is realized, it may be in the time reversal invariant VSL phase or in the time reversal broken non-Abelian pfaffian-like state from p+ip pairing. One way to distinguish them is to measure the spin gap. The time reversal breaking of the later state is quite subtle and may be revealed only through the challenging thermal Hall experiment.  We expect the cheapest anyon in our non-Abelian descendant  to be a neutral  Ising anyon with $Q=0$ and $S_z=\frac{1}{4}$. We leave to future work to design a protocal to detect and control such a neutral Ising anyon.

\section{Summary}

In summary, we studied the possible fully gapped descendants from the recently proposed vortex spin liquid (VSL)\cite{zhang2024vortex} in the context of the  experimental discovered FQSH state in twisted MoTe$_2$. We find a non-Abelian state with 12 anyons (including two independent Ising anyons) at zero magnetization and a Jain sequence of Abelian states at magnetic plateau $m=2\langle S_z\rangle=\frac{1}{p},\ p\in Z$. Therefore the VSL phase can be viewed as a powerful parent state to organize FQSH insulators at total filling $\nu_T=1$ under  continuously varying magnetizations.

\textit{Note added: } Recently there is a preprint\cite{jian2024minimal} which constructed a Z$_4$ topological order and a non-Abelian state with Ising anyons for FQSH insulator from a different perspective.  They appear to be the same topological orders as the states  from strong and weak pairing of the VSL phase in the current paper.

\textbf{Acknowledgement} YHZ thanks Chao-Ming Jian and Cenke Xu for discussions. This work was supported by the National Science Foundation under Grant No. DMR-2237031.

\bibliographystyle{apsrev4-1}
\bibliography{vsl}

\onecolumngrid
\appendix

\end{document}